\documentstyle[preprint,aps]{revtex}
\tighten
\draft
\widetext
\preprint{YITP-00-37}
\bigskip
\bigskip

\newcommand{\EQ}{\begin{equation}}
\newcommand{\EN}{\end{equation}}
\newcommand{\bea}{\begin{eqnarray}}
\newcommand{\ena}{\end{eqnarray}}
\newcommand{\eea}{\end{eqnarray}}

\newcommand{\bi}{\begin{itemize}}
\newcommand{\ei}{\end{itemize}}

\def\R{\rightarrow}

\begin{document}
\title{Localized Gravity and Higher Curvature Terms}
\medskip
\author{Olindo Corradini\footnote{E-mail: olindo@insti.physics.sunysb.edu} 
and Zurab Kakushadze\footnote{E-mail: 
zurab@insti.physics.sunysb.edu}}
\bigskip
\address{C.N. Yang Institute for Theoretical Physics\\ 
State University of New York, Stony Brook, NY 11794}

\date{September 4, 2000}
\bigskip
\medskip
\maketitle

\begin{abstract} 
{}We consider localization of gravity in 
smooth domain wall solutions of 
gravity coupled to a scalar field with a generic
potential in the presence of the Gauss-Bonnet term. 
We discuss conditions on the scalar
potential such that domain wall solutions are non-singular.
We point out that the presence of the Gauss-Bonnet term
does not allow flat
solutions with localized gravity that violate the weak energy condition. 
We also point out that
in the presence of the Gauss-Bonnet term infinite tension flat domain walls 
violate positivity. In fact, for flat solutions unitarity 
requires that on the solution the scalar potential be bounded below.  
\end{abstract}
\pacs{}

\section{Introduction}

{}In the Brane World scenario the Standard Model gauge and matter fields
are assumed to be localized on  
branes (or an intersection thereof), while gravity lives in a larger
dimensional bulk of space-time 
\cite{early,BK,polchi,witt,lyk,shif,TeV,dienes,3gen,anto,ST,BW}. The volume
of dimensions transverse to the branes is 
automatically finite if these dimensions are compact. On the other hand, 
the volume of the transverse dimensions
can be finite even if the latter are non-compact. In particular, this can be
achieved by using \cite{Gog} warped compactifications \cite{Visser} which
localize gravity on the brane. A concrete
realization of this idea was given in \cite{RS}.

{}One motivation for considering such unconventional compactifications is 
the moduli problem. In particular, the extra dimensions in such scenarios are
non-compact while their volume is finite and fixed in terms of other 
parameters in the theory such as those in the scalar potential. That is, the
expectation values of the
scalars descending from the components of the higher dimensional metric 
corresponding to the extra dimensions are actually fixed. 

{}Recently in \cite{COSM} 
one of us considered localization of gravity arising in 
$(D-1)$-dimensional (smooth)
domain wall solutions in the system of $D$-dimensional
Einstein-Hilbert gravity coupled to a single real scalar field with a generic
scalar potential. In particular, \cite{COSM} discussed conditions on the scalar
potential such that the corresponding domain wall solutions are non-singular
(in the sense that singularities do not arise at finite values of the 
coordinate transverse to the domain wall). The usual kink type of solutions
are non-singular as they interpolate between two adjacent local AdS minima
of the scalar potential. On the other hand,
there exist other non-singular solutions (subject to the aforementioned 
non-singularity conditions on the scalar potential) which do not
interpolate between AdS minima. In fact, such solutions exist even for
potentials which have no minima at all and are unbounded below. Domain walls of
this type have infinite tension. 

{}In this paper we study effects of higher curvature bulk terms on the domain
wall solutions of the aforementioned types. As was pointed out in \cite{COSM},
adding arbitrary higher curvature combinations would generically lead to 
delocalization of gravity. Moreover, if we truncate the bulk action at any
finite higher derivative level, then generic higher curvature terms would lead
to the appearance of ghosts in the Hilbert space. To avoid these difficulties,
one might consider adding special ``topological'' combinations 
which do not spoil unitarity \cite{Zwiebach,Zumino}. In this paper we focus on
the simplest non-trivial term of this type, namely, the Gauss-Bonnet 
combination\footnote{Discontinuous domain walls (that is, domain walls with
$\delta$-function-like brane sources with non-zero tension which explicitly 
break diffeomorphism invariance) in the presence of bulk Gauss-Bonnet term
were discussed in \cite{zee,mavro,neupane}. In this
paper we will not discuss such domain walls (and, therefore, we will not have 
to deal with the issues recently pointed out in \cite{zuraRS}). Rather, we will
focus on smooth domain walls (without any $\delta$-function brane sources) 
which break diffeomorphisms spontaneously.}. 
Even though it is a total derivative in four dimensions 
(in particular, it is the four-dimensional Euler invariant), it is
non-trivial in higher dimensions. Nonetheless, when expanded around a flat
Minkowski metric, the Gauss-Bonnet term does not give rise to corrections to
the propagator, so that ghost are not introduced. Another feature of the 
Gauss-Bonnet combination is that it can be supersymmetrized. 

{}Thus, even
though it is a special combination of higher curvature terms, studying domain
walls with localized gravity in the presence of the Gauss-Bonnet term gives 
some insight into the higher curvature effects on such domain walls. Thus,
for instance, we point out that the presence of the Gauss-Bonnet term
does {\em not} allow flat
solutions with localized gravity that violate the weak energy condition 
(that is, the analog of the $c$-theorem \cite{gubser}). We also point out that
in the presence of the Gauss-Bonnet term infinite tension flat domain walls 
violate positivity. In fact, for flat solutions unitarity 
requires that (on the solution) the scalar potential be bounded below. 

{}The rest of this paper, which is essentially a generalization of \cite{COSM},
is organized as follows. In section II we describe the setup within which we
will discuss solutions with localized gravity. In section III we discuss 
solutions with vanishing $(D-1)$-dimensional cosmological constant. 
Section IV contains concluding remarks.

\section{Setup}

{}In this section we discuss the setup within which we will discuss
solutions with localized gravity. Thus, consider a single real 
scalar field $\phi$
coupled to gravity with the following action\footnote{Here we focus on
the case with one scalar field for the sake of simplicity. In particular,
in this case we can absorb a (non-singular) metric $Z(\phi)$ in
the $(\nabla\phi)^2$ term by a non-linear field redefinition. This cannot
generically be done in the case of multiple scalar fields $\phi^i$, where
one must therefore also consider the metric $Z_{ij}(\phi)$.}:
\bea
S=M_P^{D-2}\int d^Dx \sqrt{-G}\biggl[ R+
 \lambda\left(R^2-4R^{MN}R_{MN}+R^{MNRS}R_{MNRS}
 \right)+\nonumber\\
 -{4\over D-2}(\nabla \phi)^2 -V(\phi)\biggr]
\label{action}
\eea 
where $M_P$ is the $D$-dimensional (reduced) Planck mass,
and the term multiplied by a free parameter $\lambda$ is the
Gauss-Bonnet combination\footnote{We are using the conventions $R^M_{NRS}=
\Gamma^M_{NR,S}-
\Gamma^M_{NS,R}+\Gamma^M_{RT}\Gamma^T_{NS}-\Gamma^M_{ST}\Gamma^T_{NR}$, 
and $R_{MN}=R^P_{MPN}$.}. 
The equations of motion read:
\begin{eqnarray}
 && {8\over{D-2}}\nabla^2\phi=V_\phi~,\\
 \label{einstein}
 &&R_{MN}-{1\over 2}G_{MN} R
 -{1\over 2}\lambda G_{MN}\left(R^2-4R^{MN}R_{MN}+R^{MNRS}R_{MNRS}
 \right)+\nonumber\\
 &&2\lambda\left(R R_{MN}-2R_{MS}R^S_N+R_{MRST}R_N^{RST}-2R^{RS}R_{MRNS}\right)
 \nonumber\\
 &&={4\over {D-2}}\left[\nabla_M\phi\nabla_N\phi
 -{1\over 2}G_{MN}(\nabla \phi)^2\right]-{1\over 2}G_{MN} V~.
\end{eqnarray}
The subscript $\phi$ in $V_\phi$ denotes derivative w.r.t. $\phi$.

{}In the following we will be interested in solutions to the above equations
of motion with the warped \cite{Visser} metric of the following form:
\begin{equation}\label{warped}
 ds^2=\exp(2A)d{\widetilde s}^2+dy^2~,
\end{equation}
where $y\equiv x^D$, the warp factor $A$, which is a function of $y$,
is independent of the coordinates
$x^\mu$, $\mu=1,\dots,D-1$, and the $(D-1)$-dimensional interval is given
by
\begin{equation}
 d{\widetilde s}^2={\widetilde g}_{\mu\nu}dx^\mu dx^\nu~,
\end{equation}  
with the $(D-1)$-dimensional metric 
${\widetilde g}_{\mu\nu}$ independent of $y$. 

{}With the above ans{\"a}tz, we have:
\begin{eqnarray}
 &&R_{\mu\nu}={\widetilde R}_{\mu\nu}-\exp(2A)\left[A^{\prime\prime}+
 (D-1)(A^\prime)^2\right] {\widetilde g}_{\mu\nu}~,\\
 &&R_{DD}=-(D-1)\left[A^{\prime\prime}+
 (A^\prime)^2\right]~,\\
 &&R_{\mu D}=0~,\\
 &&R^\mu_{\nu\rho\sigma}={\widetilde R}^\mu_{\nu\rho\sigma}+(A^\prime)^2
 \exp(2A)\left[\delta^\mu_\sigma
 {\widetilde g}_{\rho\nu}-\delta^\mu_\rho {\widetilde g}_{\nu\sigma}\right]~,\\
 &&R^D_{\nu\rho\sigma}=0~,\\
 &&R^D_{\nu D\sigma}=-\exp(2A)\left[A^{\prime\prime}+
(A^\prime)^2\right]{\widetilde g}_{\nu\sigma}~,
\end{eqnarray}
where prime denotes derivative w.r.t. $y$. Also, the $(D-1)$-dimensional 
(``tilded'') quantities such as ${\widetilde R}_{\mu\nu}$ and ${\widetilde
R}_{\mu\nu\sigma\tau}$ are calculated w.r.t. the 
$(D-1)$-dimensional metric ${\widetilde g}_{\mu\nu}$. 

{}In the following we will be 
interested in solutions where $\phi$ depends non-trivially on $y$. From the
above equations it then follows that $\phi$ is independent of $x^\mu$.  
The equations of motion for $\phi$ and $A$ then become:
\begin{eqnarray}\label{phi''}
 &&{8\over {D-2}}\left[\phi^{\prime\prime}+(D-1)A^\prime\phi^\prime\right]-
 V_\phi=0~,\\
 \label{phi'A'}
 &&(D-1)(D-2)(A^\prime)^2\left[1-(D-3)(D-4)\lambda(A^\prime)^2\right]-
 {4\over D-2}(\phi^\prime)^2+V-\nonumber\\
 &&{D-1\over D-3} {\widetilde \Lambda}\exp(-2A)\left[1-2(D-3)(D-4)\lambda
 (A^\prime)^2\right]
 -\lambda{\widetilde\chi}\exp(-4A)=0~,\\
 \label{A''}
 &&(D-2)A^{\prime\prime}\left[1-2(D-3)(D-4)\lambda(A^\prime)^2+2
 {D-4\over D-2}\lambda{\widetilde\Lambda}
 \exp(-2A)\right]+{4\over D-2}(\phi^\prime)^2+\nonumber\\
 &&{1\over D-3}{\widetilde \Lambda}\exp(-2A)
 \left[1-2(D-3)(D-4)\lambda(A^\prime)^2\right]+{2\lambda\over D-1}
 {\widetilde\chi}\exp(-4A)=0~.
\end{eqnarray}
The first equation is the dilaton equation of motion, the second equation
is the $(DD)$ component of (\ref{einstein}), 
and the third equation is a linear combination of the
latter and the $(\mu\nu)$ component of (\ref{einstein}). 
In fact, the $(\mu\nu)$ component of (\ref{einstein})
implies that ${\widetilde \Lambda}$ is a constant, and 
is nothing but the cosmological constant of the $(D-1)$-dimensional 
manifold, which is therefore an Einstein manifold, described by the metric 
${\widetilde g}_{\mu\nu}$. Our normalization of ${\widetilde \Lambda}$ is such
that the $(D-1)$-dimensional metric ${\widetilde g}_{\mu\nu}$ satisfies
Einstein's equations
\begin{equation}
 {\widetilde R}_{\mu\nu}-{1\over 2}{\widetilde g}_{\mu\nu}
 {\widetilde R}=-{1\over 2}
{\widetilde g}_{\mu\nu}{\widetilde\Lambda}~.
\end{equation}
Moreover, the quantity 
\begin{equation}\label{chi}
 {\widetilde\chi}\equiv{\widetilde R}^2-4{\widetilde R}_{\mu\nu}^2+
 {\widetilde R}_{\mu\nu\sigma\tau}^2
\end{equation} 
is also a constant (for $\lambda\not=0$)\footnote{If the corresponding 
Einstein manifold is maximally symmetric, then we have 
${\widetilde R}_{\mu\nu\rho\sigma}={\widetilde\Lambda}
\left({\widetilde g}_{\mu\rho}{\widetilde g}_{\nu\sigma}-
{\widetilde g}_{\mu\sigma}{\widetilde g}_{\nu\rho}\right)/(D-2)(D-3)$, and
${\widetilde\chi}=(D-1)(D-4){\widetilde\Lambda}^2/(D-2)(D-3)$. Generally, 
however, this Einstein manifold need not be maximally symmetric.}. 
Note that for $D-1=4$ the quantity ${\widetilde
\chi}$ is the Euler invariant for the $(D-1)$-dimensional manifold
described by the metric ${\widetilde g}_{\mu\nu}$. Finally, the aforementioned
$(D-1)$-dimensional Einstein manifold must be such that
\begin{equation}\label{condi}
 {\widetilde R}_{\mu\rho\sigma\tau}{\widetilde R}_\nu^{\rho\sigma\tau}=
 {1\over D-1}{\widetilde R}_{\alpha\beta\rho\sigma}^2 {\widetilde g}_{\mu\nu}~.
\end{equation}
This condition is automatically satisfied for maximally symmetric Einstein 
manifolds.

{}Note that we have only two fields $\phi$ and $A$, yet we have three 
equations (\ref{phi''}), (\ref{phi'A'}) and (\ref{A''}). 
However, only two of these equations are independent. This can
be seen as follows. Using the second equation one can express $\phi^\prime$ 
($A^\prime$) via $A^\prime$ ($\phi^\prime$) and $V$. One can then
compute $\phi^{\prime\prime}$ ($A^{\prime\prime}$) and plug it in the first 
(third) equation. This equation can then be seen to be automatically satisfied
as long as the third (first) equation is satisfied. As usual, this is a
consequence of Bianchi identities.

\section{Solutions with $(D-1)$-dimensional Poincar{\'e} Invariance} 

{}In this section we discuss solutions of the aforementioned equations
with ${\widetilde\Lambda}=0$ and ${\widetilde \chi}=0$. 
In this case the equations of motion read:
\bea
 \label{phi'A'1}
 &&(D-1)(D-2)(A^\prime)^2\left[1-(D-3)(D-4)\lambda (A^\prime)^2\right]
 -{4\over D-2}(\phi^\prime)^2+V=0~,\\
 \label{A''1}
 &&(D-2)A^{\prime\prime}\left[1-2(D-3)(D-4)\lambda(A^\prime)^2\right]+{4\over D-2}(\phi^\prime)^2=0~.
\eea
As in the $\lambda=0$ case, we can rewrite these equations in terms of
the following first order equations
\begin{eqnarray}
 &&\phi^\prime=\alpha W_\phi\left(1-\lambda\kappa W^2\right)~,\\
 &&A^\prime=\beta W~,
\end{eqnarray}
where 
\begin{eqnarray}
 &&\alpha\equiv\epsilon {\sqrt{D-2}\over 2}~,\\
 &&\beta\equiv-\epsilon {2\over (D-2)^{3/2}}~,\\
 &&\kappa\equiv 2(D-3)(D-4)\beta^2~,
\end{eqnarray}
and $\epsilon=\pm 1$. Moreover, the scalar potential $V$ is related
to the function $W=W(\phi)$ via
\bea
\label{potential}
 V=\left[W_\phi^2 +\eta\right]\left(1-\lambda\kappa W^2\right)^2- \eta~,
\eea
where
\begin{equation}
 \eta\equiv {(D-1)(D-2)\over 4\lambda (D-3)(D-4)}~.
\end{equation}
Note that for $\lambda>0$ the potential (\ref{potential}) is bounded 
below \cite{zee}.  
Also note that in the $\lambda\rightarrow 0$ limit from (\ref{potential}) 
we recover the familiar expression $V=W_\phi^2-\gamma^2 W^2$, where
$\gamma^2\equiv 4(D-1)/(D-2)^2$.

{}Note that (\ref{A''1}) implies the following condition:
\begin{equation}
 A^{\prime\prime}\left[1-2(D-3)(D-4)\lambda(A^\prime)^2\right]\leq 0~.
\end{equation}
It then follows that, since $A$ as well as its derivatives are
continuous, $A^{\prime\prime}$ cannot change sign. That is, we have the 
following possibilities. If $\lambda$ is negative, then we necessarily have
$A^{\prime\prime}\leq 0$. If $\lambda$ is positive, then we can have
$A^{\prime\prime}\leq 0$ subject to the following additional requirement:
$(A^\prime)^2\leq \beta^2/\lambda\kappa$. In this case gravity is 
localized as long as
$A$ goes to $-\infty$ at $y\rightarrow\pm\infty$ fast enough.  
Another possibility (for $\lambda>0$)
is that $A^{\prime\prime}\geq 0$, and $(A^\prime)^2\geq \beta^2/\lambda\kappa$.
Note that in this case gravity is {\em not} localized (as $A$ goes to
$+\infty$ at $y\rightarrow\pm\infty$). Thus, the presence of the
Gauss-Bonnet term does {\em not} allow flat 
solutions with localized gravity that
violate the weak energy condition (that is, the analog of the $c$-theorem
\cite{gubser}).

\subsection{Non-singularity Conditions}

{}In this subsection we would like to discuss the conditions on $W$ such that
the corresponding solutions do not blow up at finite values of $y$. More
precisely, in this section we will focus on solutions such that
$\phi$ is non-singular\footnote{We will refer to the corresponding domain 
walls as non-singular. However,
some of such solutions are actually singular in the sense that the
$D$-dimensional Ricci scalar $R$ 
blows up, but the singularities are located at $y=\pm\infty$
(see below).} 
at finite $y$. To begin with note that if $V$ is non-singular,
which we will assume in the following,
then $W$ and $W_\phi$ should (generically) be non-singular as well. This then
guarantees that solutions are continuous for finite values of $\phi$. However,
{\em a priori} it is still possible that $\phi$ blows up at finite values of
$y$.

{}The equation we would like to study here is 
\begin{equation}\label{phi}
 \phi^\prime=\alpha Y_\phi~,
\end{equation}
where
\bea
Y\equiv W-{\lambda\kappa\over 3} W^3~.
\label{y}
\eea
In the following we will be interested in the cases where
gravity is localized. Then the function $Y=Y(W)$ is invertible. Indeed, if
$\lambda\leq 0$, $Y(W)$ is injective for any $W$, while for 
$\lambda >0$ it is injective for $W^2\leq {1/\lambda\kappa}$, that is,
$(A^\prime)^2 \leq  {\beta^2/\lambda\kappa}$.
Thus, in these cases we can view $W$ and $V$ as functions
of $Y$. 

{}Note that (\ref{phi}) arises in a non-gravitational theory described by
the following action:
\bea\label{nongravi}
{\cal S}=\int d^D x\left[-{4\over D-2}(\partial\phi)^2-{\cal V}\right]~,
\eea
where 
\bea
{\cal V}\equiv Y_\phi^2~.
\eea
Thus, a solution of (\ref{phi}) describes a BPS solution in the theory 
(\ref{nongravi}) which depends only on $y$. The 
tension of the corresponding 
domain wall is given by
\bea
T={2\over\alpha}\left[Y(y=+\infty)-Y(y=-\infty)\right]~.
\eea
If the theory is supersymmetric, then this (up to a normalization
constant) also gives the corresponding central charge, and $Y$ is interpreted
as the superpotential.

{}Next, let us discuss the general condition for such domain walls to be 
non-singular. That is, we would like to find the condition under which
$\phi$ does not blow up at finite values of $y$. First, let as assume that
$Y_\phi$ does not vanish for any $\phi$. Then for the domain wall to be 
non-singular,
it is necessary and sufficient that the function 
\begin{equation}\label{non-sing}
 F(\phi)\equiv\int{d\phi\over Y_\phi}
\end{equation}
is unbounded at $\phi\rightarrow\pm \infty$.
That is, the non-singularity condition reads:
\begin{center}
 $Y$ should not\footnote{More precisely, this is correct up to usual
 ``logarithmic'' factors (that is, $\log(\phi)$, $\log(\log(\phi))$, {\em 
 etc.}, or, more generally, the appropriate products thereof). Thus, for 
 instance, the non-singularity condition on (\ref{non-sing}) 
 is satisfied for $Y=\xi\phi^2\log(\phi)$.} grow 
 faster than $\phi^2$ for $\phi\R\pm\infty$,
\end{center}
or, equivalently,
\begin{center}
 $W$ should not grow faster than $\phi^{2/3}$ for $\phi\R\pm\infty$.
\end{center}
On the other hand, if $Y_\phi$ vanishes at one point\footnote{Here such
a point can be at finite $\phi$ or $\phi=\pm\infty$.}, 
call it $\phi_0$, then
we have non-singular domain walls interpolating between $\phi=\phi_0$ and
$\phi=\pm\infty$ as long as at $\phi=\pm\infty$ the above non-singularity 
condition is satisfied. Finally, if $Y_\phi$ vanishes for more then one
value of $\phi$, then we have the usual non-singular domain walls of the 
kink type interpolating between the adjacent values of $\phi$ where $Y_\phi$
vanishes. Note that such domain walls have finite tension. In contrast,
non-singular solutions where $Y$ goes to $\pm\infty$ have infinite tension.
Such domain walls automatically localize gravity (provided that $Y=Y(W)$ is
invertible) as long as $W$ changes sign. 
On the other hand, for the kink type of solutions to localize gravity
it is also required that $W$ does not vanish 
at the edges of the domain wall (that is, at the points
where $Y_\phi$ vanishes).

\subsection{Positivity Conditions}

{}In this subsection we would like to discuss an additional consistency
condition on domain wall solutions in the presence of the Gauss-Bonnet
term. Thus, since we are dealing with higher curvature terms, we must make
sure that unitarity is not violated in the corresponding warped backgrounds.
As in the previous subsection, let us focus on non-singular solutions
that localize gravity.

{}Let us substitute the domain wall ans{\"a}tz (with 
${\widetilde\Lambda}={\widetilde\chi}=0$) into the action $S$
given by (\ref{action}).
We then obtain the following $(D-1)$-dimensional
action for the metric ${\widetilde g}_{\mu\nu}(x^\sigma)$:
\bea\label{action3}
 && {\widetilde S}/{\widetilde M}_P^{D-3}=\int d^{D-1}x \sqrt{-{\widetilde g}}
 \left[{\widetilde R} +
 {\widetilde\lambda}\left({\widetilde R}^2-4{\widetilde R}_{\mu\nu}^2
 +{\widetilde R}_{\mu\nu\rho\sigma}^2
 \right)\right]~,
\eea
where we have dropped the boundary terms as they vanish for non-singular domain
walls that localize gravity (for such domain walls 
$A^\prime\exp(A)$ and $A^{\prime\prime}\exp(2A)$ go to zero at 
$y\rightarrow\pm\infty$). In the last equation we are using the 
following notations:
\bea
 &&{\widetilde M}_P^{D-3}\equiv M_P^{D-2}\int dy \exp[(D-3)A]\left[1
 +2\lambda(D-3)(D-4)(A^\prime)^2\right]~,\\
\label{planck}
 &&{\widetilde \lambda}\equiv\lambda {M_P^{D-2}\over{\widetilde M}_P^{D-3}}
 \int dy \exp[(D-5)A]~.
\eea
The quantity ${\widetilde M}_P$ is interpreted as the $(D-1)$-dimensional
Planck scale, and ${\widetilde\lambda}$ is the $(D-1)$-dimensional
analog of $\lambda$. Note that in $D=5$ 
the quantity ${\widetilde\lambda}$ is infinite.
This, however, does not pose a problem as in $D-1=4$ dimensions the 
Gauss-Bonnet term is a total derivative, and if we drop the corresponding
topological term, we obtain the usual 4-dimensional Einstein-Hilbert action:
\bea
 {\widetilde S}={\widetilde M}_P^2\int d^4x \sqrt{-{\widetilde g}}
 {\widetilde R}~.
\eea
For $D>5$ the Gauss-Bonnet term is no longer a total derivative, 
and ${\widetilde\lambda}$ is finite. Note, however, that if we expand the 
Gauss-Bonnet term around the flat Minkowski solution (and this is
the $(D-1)$-dimensional background 
we must consider in accord with the original domain wall solution), it does 
not modify the graviton propagator (that is, the terms quadratic in metric
fluctuations arising from expanding the Gauss-Bonnet term combine into
a total derivative), so that unitarity is not violated \cite{Zwiebach,Zumino}.
Nonetheless, the Gauss-Bonnet term does non-trivially modify the interactions.

{}The above observation, however, is insufficient to ensure positivity. Thus,
the integrand in (\ref{planck}) is positive-definite if 
$\lambda>0$, but for $\lambda<0$ it can become
negative. This implies that ${\widetilde M}_P^{D-3}$ can in some cases be 
negative if $\lambda<0$. We would then have negative-norm states, which violate
unitarity. In fact, to ensure unitarity we should require that the
integrand in (\ref{planck}) is positive-definite for all $y$ - indeed,
\begin{equation}
 \exp[(D-3)A]\left[1+2\lambda(D-3)(D-4)(A^\prime)^2\right]
\end{equation}
is interpreted as the square of (the $y$-dependent part of) the
graviton wave-function. This then implies the following positivity 
condition:
\begin{equation}
 (A^\prime)^2\leq\beta^2/|\lambda|\kappa~,
\end{equation} 
or, equivalently,
\begin{equation}\label{positivity}
 W^2\leq 1/|\lambda|\kappa~.
\end{equation}
Note that for $\lambda>0$ this is a necessary condition for
a domain
wall to localize gravity. On the other hand, for 
$\lambda<0$ this condition ensures unitarity.

{}The above positivity condition has an important implication. Thus, it is not
difficult to see that 
infinite tension domain walls discussed in the previous subsection 
exist only for $\lambda<0$, and they violate the positivity condition
(\ref{positivity}). That is, as was already suspected in \cite{COSM}, 
infinite tension domain wall solutions are not completely consistent once
higher curvature terms are included. The reason for this is that such domain
walls are actually singular with the singularities (where the Ricci scalar
$R$ diverges) located at $y=\pm\infty$. In contrast, flat 
domain walls with finite tension are non-singular everywhere, and as long
as (\ref{positivity}) is satisfied (on the solution), they do not violate
unitarity. This implies that (for both $\lambda<0$ and $\lambda>0$) 
on the solution we have 
\begin{equation}   
 V\geq -|\eta|~,
\end{equation}
that is, the scalar potential is bounded below.

\subsection{An Example}

{}For illustrative purposes 
let us end our discussion here with a simple example of a domain wall 
with finite tension which satisfies the consistency conditions discussed
in this section. Thus, let $\lambda>0$, and let $W=\zeta\phi$ (for 
definiteness let us assume $\zeta>0$). We then
have
\begin{equation}
 Y=\zeta\phi-{\lambda\kappa\over 3}\zeta^3\phi^3~.
\end{equation}
The domain wall solution is then given by:
\begin{eqnarray}
 &&\phi(y)={1\over{\zeta\sqrt{\lambda\kappa}}}\tanh\left[\alpha\zeta^2
 \sqrt{\lambda\kappa}(y-y_0)\right]~,\\
 &&A(y)={\beta\over{\alpha\zeta^2\lambda\kappa}}\ln\left(
 \cosh\left[\alpha\zeta^2
 \sqrt{\lambda\kappa}(y-y_0)\right]\right)+A_0~,
\end{eqnarray}
where $y_0$ and $A_0$ are integration constants.

\section{Comments}

{}In this section we would like to make a few concluding remarks.
As we saw in the previous section, consistent flat domain wall solutions 
in the presence of the Gauss-Bonnet term are of the kink type, and they
interpolate between adjacent AdS minima of the scalar potential. Here we should
point out that such solutions always have consistent curved deformations
(that is, for such potentials there always exist consistent domain
wall solutions with non-vanishing $(D-1)$-dimensional cosmological
constant). 

{}As we have already mentioned in Introduction, one of the 
motivations for choosing the Gauss-Bonnet combination is that, as was 
pointed out in \cite{COSM}, generic
higher curvature terms actually delocalize gravity. Thus,
inclusion of higher derivative terms of, say, the form
\begin{equation}
 \zeta\int d^Dx \sqrt{-G} R^k
\end{equation}
into the bulk action would produce terms of the form \cite{COSM}
\begin{equation}
 \zeta \int d^{D-1}x dy\exp[(D-2k-1)A]\sqrt{-{\widetilde g}}{\widetilde R}^k~.
\end{equation}
Assuming that $A$ goes to $-\infty$ at $y \rightarrow\pm\infty$, 
for large enough $k$ the factor $\exp[(D-2k-1)A]$ 
diverges, so that at the end of the day gravity is no longer localized.
In fact, for $D=5$ delocalization of gravity takes place already at the 
four-derivative level once we include the $R^2$, $R_{MN}^2$ and $R_{MNRS}^2$
terms with generic coefficients (with the only exception being the 
Gauss-Bonnet combination).

{}A possible way around this difficulty might be that all the higher curvature
terms should come in ``topological'' combinations (corresponding to Euler
invariants such as the Gauss-Bonnet term \cite{Zwiebach,Zumino}) 
so that their presence does not
modify the $(D-1)$-dimensional propagator for the bulk graviton modes. That is,
even though such terms are multiplied by diverging powers of the warp factor,
they are still harmless. One could attempt to justify the fact that higher
curvature bulk terms must arise only in such combinations by the fact that
otherwise the bulk theory would be inconsistent to begin with due to the
presence of ghosts. However, it is not completely obvious whether it is
necessary to have only such combinations to preserve unitarity. Thus, in
a non-local theory such as string theory unitarity might be preserved,
even though at each higher derivative order there are non-unitary terms, due
to a non-trivial cancellation between an infinite tower of such terms.

{}We would like to end our discussion by pointing out that the aforementioned
difficulty with higher curvature terms does not arise in theories with 
infinite-volume non-compact extra dimensions 
\cite{GRS,CEH,DGP0,witten,DVALI,zura,DG}. 
However, in such scenarios consistency of the coupling between bulk 
gravity and brane matter might give rise to additional constraints. Thus, 
in some cases the brane world-volume theory must be
conformal \cite{zura}. 
In such cases it would be interesting to understand if there is a
relation to \cite{BKV}. 

\acknowledgments

{}This work was supported in part by the National Science Foundation.
Z.K. would like to thank Albert and Ribena Yu for financial support.

\end{document}